\documentclass[12pt]{article}
\pdfoutput=1

\usepackage{graphicx}
\usepackage{psfrag}
\usepackage{enumerate}
\usepackage{soul}

\usepackage{subfig}
\usepackage{jheppubcustom}
\usepackage{latexsym}
\usepackage{natbib}
\usepackage{dsfont}

\newcommand{\be}{\begin{equation}}
\newcommand{\ee}{\end{equation}}
\renewcommand{\Im}{{\rm Im}}

\def\[{\left [}
\def\]{\right ]}
\def\({\left (}
\def\){\right )}

\def\r2{\sqrt{2}}

\def\O{{\mathcal O}}

 \def\simleq{\; \raise0.3ex\hbox{$<$\kern-0.75em
      \raise-1.1ex\hbox{$\sim$}}\; }
   \def\simgeq{\; \raise0.3ex\hbox{$>$\kern-0.75em
      \raise-1.1ex\hbox{$\sim$}}\; }

\DeclareMathOperator{\sech}{sech}

\newcommand{\de}{\partial}

\newcommand{\del}{\delta}

\newcommand{\lam}{\lambda}

\newcommand{\sig}{\sigma}

\newcommand{\om}{\omega}

\renewcommand{\k}{\vec{k}}
\newcommand{\x}{\vec{x}}
\newcommand{\avg}[1]{\langle #1 \rangle}
\newcommand{\ra}{\rangle}
\newcommand{\la}{\langle}

\newcommand{\bbibitem}[1]{\bibitem{#1}\marginpar{#1}}

\newcommand{\figref}[1]{Fig.~\ref{#1}}
\newcommand{\secref}[1]{Sec.~\ref{#1}}

\newcommand{\appref}[1]{Appendix \ref{#1}}

\newcommand{\bi}{\begin{itemize}}
\newcommand{\ei}{\end{itemize}}

\def\Label#1{\label{#1}%
  \smash{\hbox to0pt{\raise1ex\hbox{\tiny[#1]}\hss}}}
\def\noLabels{\let\Label=\label}
\def\nobbibitem{\let\bbibitem=\bibitem}

\begin{document}

\noLabels
\nobbibitem

\DeclareGraphicsExtensions{.pdf,.png,.gif,.jpg,.eps}

\title{D-brane scattering and annihilation}
\author{Guido D'Amico, Roberto Gobbetti, Matthew Kleban, and Marjorie Schillo}

\emailAdd{gda2@nyu.edu, rg1509@nyu.edu, mk161@nyu.edu, mls604@nyu.edu}

\affiliation{\it Center for Cosmology and Particle Physics\\
New York University  }

\abstract{
We study the dynamics of parallel brane-brane and brane-antibrane scattering in string theory in flat spacetime, focusing on the pair production of  open strings that stretch between the branes.  We are particularly interested in the case of scattering at small impact parameter $b < l_{s}$, where there is a tachyon in the spectrum when a brane and an antibrane approach within a string length.  Our conclusion is that despite the tachyon, branes and antibranes can pass through each other with only a very small probability of annihilating, so long as $g_{s}$ is small and the relative velocity $v$ is neither too small nor too close to 1.   Our analysis is relevant also to the case of charged open string production in world-volume electric fields, and we make use of this T-dual scenario in our analysis.  We briefly discuss the application of our results to a stringy model of inflation involving moving branes.}

\maketitle

\begin{center}
\begin{table}[t]
\begin{center}
\centering
\begin{tabular}{| l | l |}
	\hline
	\multicolumn{2}{|c|}{\bf  \it Dramatis Person\ae} \\
	\hline 
	$ \sigma \in [0, \pi],\, \tau$  & string worldsheet coordinates \\
		\hline
		$ \alpha'=1/2$, $T_{0} \equiv 1/(2 \pi \alpha')=1/\pi$ & string scale and tension in our units \\
		\hline
	$v_0, v_\pi; \,\,\, e_0, e_\pi$ & velocities of the branes; charges of the string ends \\
	\hline
	$\gamma = \(1-v_0^2\)^{-1/2}=\(1-v_\pi^2\)^{-1/2}$ & Lorentz factor, used in center of mass frame $v_0 = -v_\pi$ \\
	\hline
	$\chi = \frac{1}{\pi} \left| \tanh^{-1}\(v_\pi\) - \tanh^{-1}\(v_0\) \right|$  & relative rapidity times $\pi^{-1}$ in the brane scattering frame, \\
	 $\chi =  \frac{1}{\pi} \left| \tanh^{-1}\(\pi e_0 E\) + \tanh^{-1}\(\pi e_\pi E\) \right|$& goes to infinity at the critical field in the electric frame \\
		\hline
	$D;\,\,\, \la n \ra$ & degeneracy of states;  particle or string number density\\
	\hline
	$b;\,\,\, p+1$ & impact parameter; $D_p$-brane worldvolume dimension \\
	\hline
	$2 \Im\({\cal A}\) = - \ln P_{vac}$ & $\cal A$ is the vacuum-vacuum amplitude, 	 $P_{vac}$  the vacuum \\ & persistence probability (prob. of producing nothing) \\
	 \hline
	$l_{*}$  & stopping distance (c.f.~intro to \secref{bscat}) \\
	 \hline
 \end{tabular}
\end{center}
\end{table}
 \end{center}

\section{Introduction}
In string theory, D-branes  \cite{Shenker:1995xq, Polchinski:1995mt} are fundamental objects at the same level as strings themselves.  They play a central role in non-perturbative string dualities, are essential ingredients in  string phenomenology, and string theory realizations inflationary or de Sitter spacetimes often involve branes and antibranes.   However, compared to strings (or particles) we have a  poor understanding of D-brane dynamics.  There is  no systematic  approach, and the study of brane/brane scattering has been mostly restricted to situations in which some trick or special symmetry can be utilized -- for instance nearly supersymmetric situations such as parallel brane-brane scattering at low velocities (with some notable exceptions, for instance \cite{Bachas:1995kx, Douglas:1996yp, McAllister:2004gd}).   To our knowledge there has been little or no study of brane-antibrane scattering.  In this work we will take a few steps in that direction. 

\begin{figure}
\begin{center} 
 \includegraphics[width=0.7\textwidth]{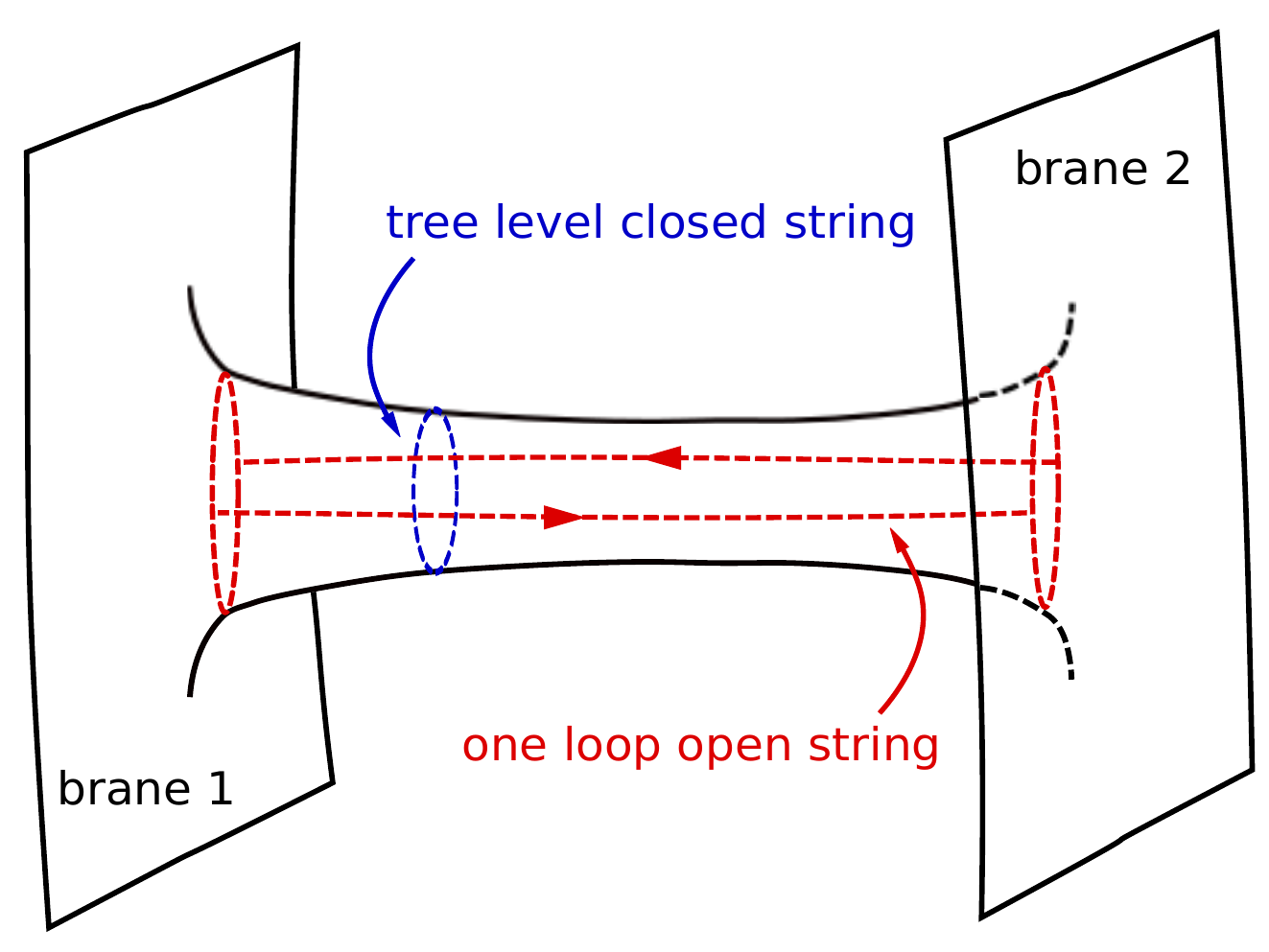}
\end{center}
\caption{\label{annulusfig}The annulus diagram can either be interpreted as the tree level exchange of a closed-string, or a 1-loop open string vacuum diagram.  Cutting the diagram along the oriented red dashed lines shows a pair of stretched open strings produced as the branes scatter. }
\end{figure}

Pairs of D-branes can interact through open strings that begin on one brane and end on the other, or closed strings they emit/absorb.  In string perturbation theory, the leading diagram contributing to this interaction is the annulus \figref{annulusfig}, which can be interpreted either as the tree-level exchange of a closed string or as a 1-loop diagram describing open strings stretching between the branes.  
If the two branes are in relative motion (or for the case of brane-antibrane in close proximity \cite{Banks:1995ch}) this diagram has an imaginary part that computes the rate of pair production of open string states that stretch between the branes \cite{Bachas:1995kx}.  In this work we will focus on the annulus amplitude, and not consider higher-order in $g_{s}$ processes such as closed string Brehmsstralung.

\begin{figure}
\begin{center} 
 \includegraphics[width=.75 \textwidth]{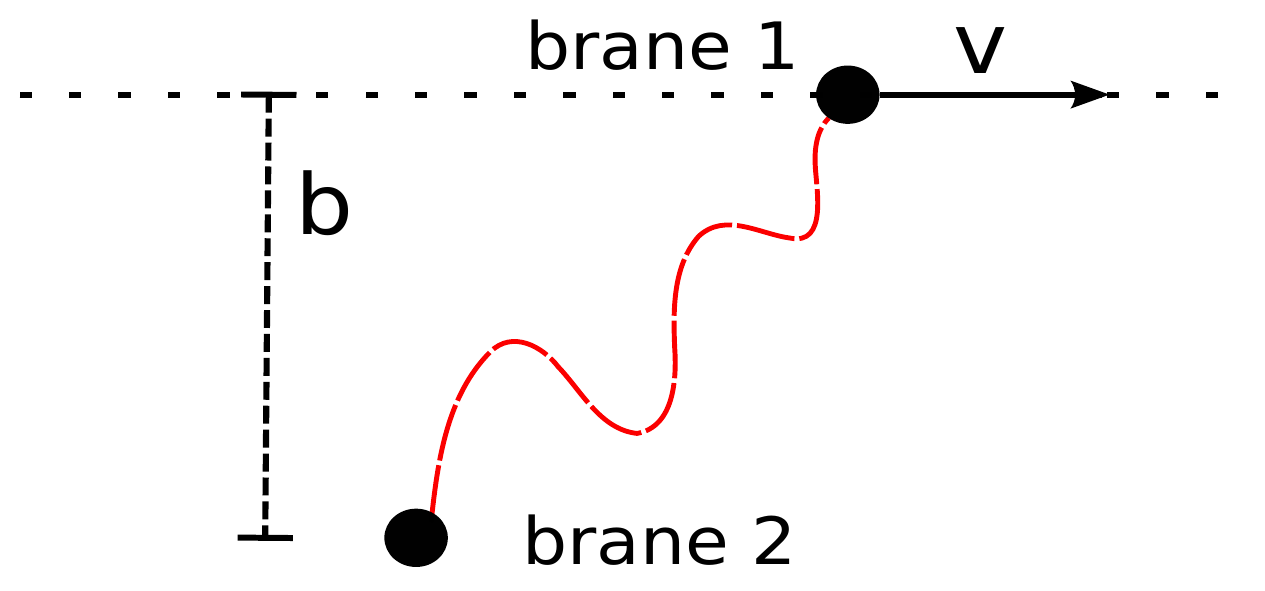}
\end{center}
\caption{\label{scatfig}Brane scattering with impact parameter $b$, and a stretched open string. }
\end{figure}

The physical reason for open string production is that the masses of strings stretched between the branes change with time as the branes approach and then recede from each other.  At small $g_{s}$,  the spectrum of  string states for a static string is  schematically 
$$m^2 \sim j + l^2,$$ 
where $l$ is proportional to the length of the string and $j$ is an integer corresponding to the excitation mode of the string.  Hence, for one stationary brane and one moving with constant velocity $v$ (so that $v_0=0,v_\pi=v$; see \figref{scatfig}) one expects the mass of stretched strings to obey a formula like
\be \label{mnaive}
m^{2}_{\rm naive} \sim j + b^{2} + (v t)^{2}.
\ee
This raises two interesting questions, which much of the paper is devoted to answering:  
\bi

\item Since $v<1$, the naive formula \eqref{mnaive} implies that string production should be exponentially suppressed for string states with  $j \gg 1$ even if $b = 0$, and  for all $j$ if $b \gg 1$.  This follows from the fact that the non-adiabaticity parameter $\dot m/m^{2} < v/(b^{2}+j)$ for all times $t$.    Concretely, one expects 
\be \label{Gnaive}
\langle n \rangle_{\rm naive} \sim e^{-(b^{2}+j)/v},
\ee
where $\la n \ra$ is the number density at level $j$.
 Instead, the annulus diagram (and results from open string field theory) imply that
\be \label{st} 
  \la n \ra_{\rm string \, \, theory} \sim e^{- (b^{2}+j)/\pi \chi},
 \ee
where $\pi \chi \equiv | \tanh^{-1} v |$ is the rapidity.  Hence, for $b^{2} + j \simleq \pi \chi$ the production is unsuppressed.  Why is this, and what are the implications?

\item For brane-antibrane the lowest value of $j$ is $-1$,  so there is a tachyon when $b<1$ and $t = 0$.  This tachyon can condense, which if $v=0$ leads to brane-antibrane annihilation.  What happens when $v>0$?

\ei

\paragraph{Tachyon condensation:} Our answer to the second question is as follows.  If a  field with $m^2$ initially greater than zero becomes tachyonic, its wavefunction spreads by an amount that depends on how long it remains in the tachyonic regime.  This is under analytical control so long as non-linear corrections to the quadratic action do not become important, and if the field eventually acquires $m^2>0$ again, this spread simply corresponds to a finite amount of particle production.  In string theory at small $g_{s}$, interactions of the canonically normalized tachyon field are suppressed by powers of $g_{s}$, meaning the wavefunction must spread a distance $\sim 1/\sqrt{g_{s}}$ before the quadratic approximation breaks down.  Since the time the mode remains tachyonic is $t \sim 1/v$, for  $g_{s} \ll 1$ there is a velocity $v \ll 1$ above which the field is very unlikely to have time to reach the non-linear regime.  Furthermore the kinetic energy of the branes scales as $1/g_{s}$.  Hence while the $j=-1$ mode of the stretched strings is always produced copiously, at weak coupling this has a small effect on the motion of the branes, and annihilation is very improbable.  Interestingly, this conclusion becomes incorrect at ultrarelativistic velocities, for reasons related to the first question raised above.

\paragraph{Enhanced string production:} The formula \eqref{st} has a very interesting consequence.  Because the density of open string modes grows exponentially with $\sqrt{j}$, when the scattering is ultra-relativistic ($\chi \sim \ln \gamma \gg 1$) there is an exponentially large amount of open string production.\footnote{In this regime at finite $g_s$ there can be a large amount of closed string Bremsstrahlung \cite{Bachlechner:2013fja} which may dominate the open string production, and  one must be careful that the force exerted by this radiation does not substantially alter the velocities of the branes and invalidate the constant-$v$ approximation used to derive these results.}  For scattering at in the moderately relativistic regime $\chi \sim 1$, the production of open strings occurs primarily in the lowest few modes.

The energy to produce strings comes from the brane kinetic energy.  Since the number of produced strings can be very large, in the ultrarelativistic regime this can cause the branes to stop very suddenly.  In order to calculate this stopping distance, one needs the number density of produced strings and the energy per string.  The enhancement in the production rate \eqref{st} relative to \eqref{Gnaive} is an intrinsically stringy phenomenon that in string field theory manifests itself in a modification of $m_0^2$, but not  of the energy in a long stretched string $\sim vt$ (so that contrary to the proposal of  \cite{McAllister:2004gd}, the force the stretched strings exert on the branes at large separations is \emph{not} velocity-dependent).  We investigate this by direct calculation of the classical string energy,  by  Euclidean instanton methods, and from the equations of motion of the open string field theory describing charged open strings in a background electric field.  With this formula in hand we compute the stopping distance, and confirm that the qualitative conclusion of \cite{McAllister:2004gd} is indeed correct -- the stopping distance in ultra-relativistic brane-brane scattering \emph{decreases} with increasing brane velocity.  Interestingly, for brane-antibrane scattering and for $p$-brane-$p$-brane scattering with $p \leq 4$ the behavior is more complex:  the stopping distance \emph{increases} with velocity in the non-relativistic regime, and then turns over and decreases at higher velocities (\figref{stopdist}, \figref{stopdistp}).
 
This dynamics is of particular interest in brane inflation models.  For instance, in unwinding inflation \cite{D'Amico:2012ji}, string production alters the classical trajectory of the inflaton (which is the distance between two branes moving around a compact space and repeatedly passing close to each other) by providing an additional force.  It also provides a source for inflaton fluctuations, since the production of a string is a local process.  Perhaps most importantly, brane-antibrane annihilation ends inflation and the resulting radiation reheats the universe.

The phenomenon of enhanced massive mode string production also has a fundamental implication for  brane world-volume electric fields.  Branes in relative motion are T-dual to branes with a nonzero electric field, and strings stretched between the moving branes map to strings with a non-zero net charge under the dual field.  The statement that neither brane's velocity can exceed the speed of light translates into the existence of a maximum value for the electric field, $E_{crit} = 1/(2\pi \alpha' | {\rm max} \, e_i |)$, where $e_i$ are the charges at the ends of the string.
In this electric frame, charged open strings are produced on the branes, in the string theoretic analogue of Schwinger's classic result~\cite{Schwinger:1951nm} for electron-positron pair production in an electric field \cite{Bachas:1992bh}.  The key difference with naive field theory is again the factor of $\chi$ in \eqref{st}, together with the exponentially growing density of states.  In standard field theories like QED, super-Schwinger electric fields $E > m^{2}/e$ are possible and physical, and the rate of discharge due to charged particle production is finite (for example, an $E > 0$ initial state is perfectly well-behaved in the 1+1 dimensional \emph{massless} Schwinger model).  But for the reasons described just above in the context of brane scattering, the closer the field comes to the critical value, the \emph{less} time it takes to discharge it to zero, consistent with the hypothesis that  \eqref{st}  prevents $E$ from  exceeding $E_{crit}$ even temporarily.  

\paragraph{Organization:} The paper is organized as follows.  In \secref{anndig} we begin by expanding the imaginary part of the annulus diagram in an appropriate limit, noting the differences between brane-brane and brane-antibrane, and between brane scattering and its T-dual -- charged strings in a worldvolume electric field.  In \secref{prodrate} we review the relation between the imaginary part of the vacuum loop diagram and the number density of produced particles in field theory, and apply the results to string theory.  In \secref{explain chi} and \secref{massimo} we derive the stringy $e^{m^2/\chi}$ dependence in three different ways:  from the annulus, from a string instanton, and using the equations of motion of string field theory in a background electric field.  In \secref{bscat} we consider the dynamics of brane-brane and brane-antibrane scattering.  \secref{sl} computes the ``stopping length" -- the distance the branes recede before the energy in stretched strings equals their initial kinetic energy -- while \secref{ann} computes the probability for brane-antibrane annihilation.  In \secref{uwi} we discuss the application of our results to unwinding inflation, and then conclude in \secref{conc} with a list of open questions.

\section{The annulus diagram} \label{anndig}

Consider two parallel Dirichlet $p$-branes moving with constant relative velocity $\vec{v}$ in flat spacetime (\figref{scatfig}). The branes  approach each other until they reach some minimal distance $b$, and then recede.  Since D-branes have a mass that scales as $\sim 1/g_s$, in the limit $g_s \to 0$ any acceleration due to string production or radiation should be small and the approximation of constant $v$ valid, at least for short time-scales.

T-duality in the $\hat{v}$ direction maps this scenario to a pair of $p+1$ branes a distance $b$ apart that are extended in the $\hat v$ direction with a constant world-volume electric field pointing in that direction. The electric field maps to the velocity via 
$$\vec{v}_0 +\vec{v}_\pi \leftrightarrow 2\pi \alpha' (e_\pi - e_0)\vec{E}.$$
The notation here refers to a string parametrized by $\sig \in (0,\pi)$, so that $\vec v_{0(\pi)}$ (correspondingly, $e_{0(\pi)}$) are the velocity (respectively, charge) at the boundaries $\sig = 0(\pi)$~\cite{Bachas:1995kx}.  The relative velocity between the branes $\vec v = \vec v_0 - \vec v_\pi$, and we can also choose a frame in which all velocities are parallel and transverse to the brane worldvolume.  Again, in the limit $g_s \to 0$ the discharge of the electric field by charged string production will be slow, and the approximation of constant $\vec E$ should be valid.

The imaginary part of the annulus amplitude is related to the probability of producing string pairs that stretch between the branes (the real part, which we will not be interested in, computes the force between the branes).  
For the case of two parallel $p$-branes -- either charged strings in a world-volume electric field or brane scattering -- the imaginary part of the annulus amplitude is  \cite{Burgess:1986dw, Bachas:1992bh, Bachas:1995kx}:
\be
\label{eq:annulus}
	\Im[\mathcal{A}] =  \frac{C}{4} \(\frac{L}{2\pi} \)^\lambda \chi^{\lambda/2} \sum_{r=1}^\infty \frac{1}{r^{(\lambda+2)/2}} \exp\( \frac{-r \pi m_0^2}{\chi} \) 2 \left\{  Z_{F}\(i \frac{r}{\chi}\) - (-1)^r  Z_B\(i \frac{r}{\chi}\)\right\}
\ee
where we use units in which $\alpha'=1/2$ and the string tension $T_0 = 1/(2 \pi \alpha')=1/\pi$, $m_0 = T_0 b=b/\pi$ is the string mass corresponding to the minimal distance between the branes $b$, $L$ is the side length of the box in which we quantize momenta (\emph{i.e.}~$L^{p}$ is the volume of the brane),  and $Z_{F(B)}$ is the  fermion (boson) string partition function.   Lastly, 
\be
\begin{split}
	C &=   \begin{cases} \frac{T L | E(e_0 + e_\pi)|}{2\pi} \quad & \text{Electric} \\ 1 & \text{Scattering} \end{cases} \, 
	\qquad \quad
	\lambda =  \begin{cases} p-1  \quad & \text{Electric} \\ p & \text{Scattering} \end{cases}  \,  \\
	\chi &=  \begin{cases}\left|  \frac{1}{\pi}\(\tanh^{-1}\(\pi e_0 E\) + \tanh^{-1}\(\pi e_\pi E\)\) \right|
	\quad & \text{Electric} \\  \left| \frac{1}{\pi}\(\tanh^{-1}\(v_\pi\) - \tanh^{-1}\(v_0\)\) \right|& \text{Scattering} 	\end{cases} 
\end{split}
\ee
where $T$ is a time interval.  Due to supersymmetry, $Z_F(i \frac{r}{\chi}) = Z_B(i \frac{r}{\chi}) ={1 \over 2} \Theta_2(i \frac{r}{\chi})^4 \eta(i \frac{r}{\chi})^{-12}$, where $\Theta$ and $\eta$ are the theta function and Dedekind $\eta$ functions respectively (we use the conventions defined in \cite{Polchinski:1998rq}).  The factor of 2 multiplying the curly braces in \eqref{eq:annulus} arises because one can interchange the ends of the string.

It is important to emphasize that this result is \emph{exact} in $\alpha'$ and  $m_{0}^{2}/\chi$, so long as $\vec v, E$ is constant.  (There are of course corrections at higher order in $g_{s}$ and from terms proportional to derivatives of $\vec v, E$.)

The factor  of the time-interval $T$ arises in the electric case because the production of charged states is a continuous process that occurs at a constant rate, at least so long as the electric field remains constant and non-zero.  By contrast in brane scattering strings are produced mostly during the interval when the branes are close together and the string mass is near its minimum, and the total number produced (in a single scattering event) is finite.\footnote{Given this, one might wonder how the two results can be T-dual.  To see the answer, note that \eqref{eq:annulus} is the result for a single brane-brane scattering event in non-compact space.  To perform a T-duality the $\hat{v}$ direction should instead be a circle, say of circumference $L$.  Then in a time $T$, $T|v_0-v_\pi|/L$ scattering events will occur.  T-dualizing this factor and equating $|v_0 - v_\pi|$ with $\pi |E(e_0+e_\pi)|$, one finds $T|v_0-v_\pi|/L \leftrightarrow TL'E(e_0+e_\pi)/(2\pi) = C$.  }
   
 Expanding the partition functions in $ \exp\(-2\pi r/\chi\)  \ll 1$, we can re-write eq. \eqref{eq:annulus} as:
 \be \label{BBannulus}
 \begin{split}
	 \Im[\mathcal{A}] =&  \frac{C}{4} \(\frac{L}{2\pi} \)^\lambda \chi^{\lambda/2} \sum_{r=1,3,5}^\infty \frac{1}{r^{(\lambda+2)/2}} \exp\( \frac{-r \pi m_0^2}{\chi} \) \\
 	& \times \left\{ 32 + 512 \exp\(-\frac{2\pi r}{\chi}\) + 4608\exp\(-\frac{4\pi r}{\chi}\) + \cdots \right\} \,.
\end{split}
\ee

One can interpret \eqref{BBannulus} by looking at its field theoretic counterpart.  Schwinger's classic result for the imaginary part of the vacuum amplitude for a charged particle in a background, constant electric field $E$ in $d$ spacetime dimensions is:
 \be \label{schwing}
	 \Im[{\cal A}_{\rm field \, theory}] = \frac{D}{4}  T \(\frac{L}{2 \pi}\)^{d-1}  (eE)^{\frac{d}{2}}  \sum_{r=1}^\infty \frac{(-1)^{(r+1) (2 S+1)}}{r^{d/2}}\exp\left( \frac{-r \pi m^2} {eE} \right) \, ,
 \ee
where $S$ is the spin of the produced particle, $e, m$ its charge, mass, $D$ the number of  degrees of freedom, and $T$ is again a time interval.    To compare this to string theory, set $p=d-1$, consider the weak field limit $\chi \approx e E$, and note 
 that the quantity in curly braces in \eqref{BBannulus} can be re-written as 
$$
\sum_{r=1,2,3,4,\cdots}^\infty \left\{16 (-1)^{r+1} + 16  + 256 (-1)^{r+1}  \exp\(-\frac{2\pi r}{\chi}\) +256 \exp\(-\frac{2\pi r}{\chi}\) +  \cdots \right\} \,.
$$
The integers are the degeneracies of open string states stretched between two D-branes, with the appropriate factors of $(-1)^{r+1}$ and $e^{-r  m_{j}^{2}/\chi}$ corresponding to their spin and mass at level $j$.
Thus, the only  apparent difference between Schwinger's result in field theory and the string theory annulus amplitude is at strong fields where $\pi \chi \sim \tanh^{-1} e E$ differs from $e E$.

In the case of a parallel brane-antibrane pair, supersymmetry is broken and the partition functions are altered in the simple way described in \cite{D'Amico:2012ji}, resulting in:
\be \label{BABannulus}
\begin{split}
	\Im[\mathcal{A}] =&  - \frac{C}{4} \(\frac{L}{2\pi} \)^\lambda \chi^{\lambda/2} \sum_{r=1}^\infty \frac{1}{r^{(\lambda+2)/2}} \exp\( \frac{-r \pi m_0^2}{\chi} \)  \\
	&\times \eta\(i\frac{r}{\chi}\)^{-12}\left\{ (-1 + (-1)^r)\Theta_3\(i\frac{r}{\chi}\)^4 + (1+(-1)^r)\Theta_4\(i\frac{r}{\chi}\)^4\right\} \\ 
	=&   \frac{C}{4} \(\frac{L}{2\pi} \)^\lambda \chi^{\lambda/2} \sum_{r=1}^\infty \frac{1}{r^{(\lambda+2)/2}} \exp\( \frac{-r \pi m_0^2}{\chi} \) \\
	& \times \left\{ 2(-1)^{r+1} \exp\(\frac{\pi r}{\chi}\) + 16 +72(-1)^{r+1} \exp\(\frac{-\pi r}{\chi}\) +256 \exp\(\frac{-2\pi r}{\chi}\) + \cdots\right\} \, .
\end{split}
\ee
Again, this takes the form of a sum over the string states, with the integer coefficients being the multiplicities of the stretched open string states.  The first term corresponds to the  tachyon (a complex boson, hence the factor of $ 2(-1)^{r+1}$), the next term are the 16 massless fermions, etc.


\subsection{The rate of open string production} \label{prodrate}
The rate of string or particle production is determined by the imaginary part of the  effective action \eqref{BBannulus} or \eqref{BABannulus}, but the precise relation  has sometimes been misunderstood in the literature (both in string and field theory).  In field theory, the number density of produced pairs of particles is given by the \emph{first term} in the sum over $r$, \emph{not} the entire sum (a very clear discussion in the context of the Schwinger effect can be found in \cite{Holstein:1999ta}).  This difference between the sum and the first term is crucial for tachyonic or massless fields, where the terms in the sum are unsuppressed or  exponentially increasing with increasing $r$.  Before considering string production, we review these facts for a field with a time-dependent mass.    Recent work that considered related issues in string theory is \cite{Silverstein:2014yza}.


Consider a free field   that satisfies the time-dependent equation of motion
\be \label{tmass}
	\Box \psi + m(t)^2 \psi = 0, \,\,\,\,\,\,m(t)^2 = m_0^2 + A^2t^2 \, ,
\ee
where $m_0$ and $A$ are constants.  
The imaginary part of the effective action in $d$ space-time dimensions for $D$ bosonic or fermionic degrees of freedom is (c.f. \appref{app:scalars})
\be \label{VacP}
\begin{split}
	-2 \Im[{\cal A}_{\rm field \, theory}] = - \frac{D}{2} \(\frac{L}{2 \pi}\)^{d-1} A^{\frac{d-1}{2}}  \sum_{r=1}^\infty \frac{(-1)^{(r+1) (2 S+1)}}{r^{(d+1)/2}}\exp\left( \frac{-r \pi m_0^2}{A} \right) 
	= \ln(P_{vac}) ,
\end{split}
\ee
where  $S=0,1/2$ for bosons, fermions.
 The last equality follows because the norm-square of the vacuum-to-vacuum transition amplitude is the vacuum persistence probability  $P_{vac}$ (the probability of producing zero particles after infinite time, given that the initial state was the vacuum).  

For a Poisson process, the probability of zero events is $ e^{-\langle N \rangle}$.  If pair production were a Poisson process (i.e. if each pair production event was independent of any others), one would have $\langle n_{\rm pair} \rangle =  \la n \rangle/2 = - \ln(P_{vac})  = 2  \Im[{\cal A}]$.  However, pair production cannot be a Poisson process.
This is most obvious for a fermionic field, where the statistics prohibits the production of more than one pair in the same state.   In fact in the model \eqref{tmass} we can explicitly compute both $P_{vac}$ and $\langle n \rangle$ (see \appref{app:scalars}).  
As we will now show, the result is very simple: for both bosons and fermions $\langle n \rangle/2 =  \la n_{\rm pair} \ra$ is equal to the \emph{ first term} in the sum over $r$ in \eqref{VacP}. 

\paragraph{Fermions:}
Due to Fermi statistics, for each wavenumber $\vec k$  we can either produce zero or one pair of particles with zero total momentum.  The expected number of particles with wave number $\vec k$ and spin $\lambda$ is:
\be
	\langle n_{\vec k, \lambda} \rangle = 0 \times P_0(\vec k) + 1\times P_0(\vec k) \omega_k,
\ee
where $P_0(\vec k)$ is the probability of producing zero particles with wavenumber $\vec k$ and spin $\lambda$, and $\omega_k \equiv P_1(\vec k)/P_0(\vec k)$, so that $P_0(\vec k) + P_0(\vec  k) \omega_k = 1$.
Therefore
\be \label{P0 of nk fermion}
	P_0(\vec k) = \frac{1}{1+ \omega_k} = 1-\langle n_{\vec k,\lambda} \rangle = 1 - e^{-\pi(k^2+m_0^2)/A} \, ,
\ee
where in the last equality we have used \eqref{nk} for $\langle n_{\vec k} \rangle$, which we compute using standard Bogolubov methods in \appref{app:scalars}.  

The overall number of particles in volume $L^{d-1}$ is therefore 
\be \label{n}
	\langle n \rangle = \( \frac{L}{2 \pi} \)^{d-1} \sum_\lambda  \int d^{d-1} k \, e^{-\pi(k^2+m_0^2)/A} =  D \left( \frac{L}{2\pi} \right)^{d-1} A^{(d-1)/2} \exp\left(\frac{-\pi m_0^2}{A}\right) \, ,
\ee
which is twice the first term in the sum for $2\Im[\cal A]$ \eqref{VacP}.
To check the consistency of this result with \eqref{VacP}, note that the vacuum persistence probability $P_{vac}$ is the probability of producing zero pairs of any wavenumber $\vec k$ and any spin $\lambda$.  Therefore:
\be
\begin{split}
\label{Pvac}
	 \ln(P_{vac}) &= {1 \over 2} \ln \prod_{\k, \lambda} P_0(\vec k) = {D \over 2}\(\frac{L}{2\pi}\)^{d-1} \int d^{d-1}k \, \ln(P_0(k)) \\
	&= -  {D \over 2} \(\frac{L}{2\pi}\)^{d-1} \frac{2 \pi^{\frac{d-1}{2}}}{\Gamma(\frac{d-1}{2})} \int dk \, k^{d-2} \sum_{r=1}^\infty \frac{1}{r} \exp\left(-r \pi (k^2+m_0^2)/A\right) \\ 
	&= - {D \over 2} \(\frac{L}{2\pi}\)^{d-1}  A^{\frac{d-1}{2}}  \sum_{r=1}^\infty \frac{1}{r^{(d+1)/2}} \exp\left(\frac{-\pi r m_0^2}{A}\right) \, ,
\end{split}
\ee
in agreement with  \eqref{VacP}.  The factor of $1/2$ in the first equality arises from momentum conservation -- the number of produced particles of momentum $\vec k$ equals the number with momentum $- \vec k$, and hence the product over all $\vec k$ is a double counting.

\paragraph{Bosons:}
The case of bosonic modes is slightly more complicated, since Bose-Einstein statistics allow for multiple pairs to be produced -- but the statistics are still not classical, the process is not Poisson, and the expected number of pairs  again turns out to be simply the first term in the sum in \eqref{VacP}.
It turns out (see \appref{app:scalars}) that
\be
	\frac{P_j(\vec k)}{P_{j-1}(\vec k)} = \om_k, 
\ee
independent of $j$.  Thus we can write the expected number of particles with wavenumber $\vec k$ as:
\be 
\begin{split}
	\langle n_{\vec k, \lambda} \rangle &= 0\times P_0(\k) + 1 \times P_0(\k)\omega_k  +  2\times P_0(\k)\omega_k^2 + 3\times P_0(\k)\omega_k^3 + \dots \\
	&= \frac{P_0(\k)\omega_k}{(1-\omega_k)^2}.
\end{split}
\ee
The total probability sums to one:
\be
	1 = P_0(\k)(1+\omega_k +\omega_k^2 +\omega_k^3 +\dots)
\ee
and therefore,
\be
	P_0(\k) = 1-\omega_k = \frac{1}{1+\langle n_{\k,\lambda} \rangle}.
\ee
Again, we can verify the statement that $\langle n \rangle/2$ is simply the first term in \eqref{VacP} by using the above relations to calculate the vacuum persistence probability:
\be
\begin{split}
	\ln(P_{vac}) &= -{D \over 2} \( \frac{L}{2\pi} \)^{d-1} \int d^{d-1}k \, \ln(1+\langle n_{k,\lambda} \rangle) \\
	&=- {D \over 2} \(\frac{L}{2\pi}\)^{d-1} \frac{2 \pi^{\frac{d-1}{2}}}{\Gamma(\frac{d-1}{2})}  \int dk \, k^{d-2} \sum_{r=1}^\infty \frac{(-1)^{r+1}}{r} \exp\left(-r \pi (k^2+m_0^2)/A\right) \\ 
	&=- {D \over 2} \(\frac{L}{2\pi}\)^{d-1}  A^{\frac{d-1}{2}}   \sum_{r=1}^\infty \frac{(-1)^{r+1}}{r^{(d+1)/2}} \exp\left(\frac{-\pi r m_0^2}{A}\right) .
\end{split}
\ee

\paragraph{String theory:}
The imaginary part of the annulus amplitude computes $P_{vac}$ in string theory.
Because it follows  from the quantum statistics in free field theory, the analysis above should apply to the string modes at weak coupling.     Therefore we conclude that the expected number density of produced strings is simply two times the first term in the sum over $r$ in \eqref{BBannulus} for the brane-brane case and in \eqref{BABannulus} for the brane-antibrane case, i.e.:
\be \label{BBnumden}
\begin{split}
	\langle n \rangle_{DD} =&   \(\frac{1}{2\pi} \)^p \chi^{p/2} \exp\( \frac{- b^2}{\pi\chi} \) \\
	& \times \left\{ 32 + 512 \exp\(-\frac{2\pi}{\chi}\) + 4608\exp\(-\frac{4\pi}{\chi}\) + \cdots \right\} \, \\
	\langle n \rangle_{D\bar{D}} =&   \(\frac{1}{2\pi} \)^p \chi^{p/2}\exp\( \frac{- b^2}{\pi\chi} \) \\
	& \times \left\{ 2 \exp\(\frac{\pi}{\chi}\) + 16 + 72 \exp\(\frac{-\pi}{\chi}\) +256 \exp\(\frac{-2\pi }{\chi}\) + \cdots\right\} \, ,
\end{split}
\ee
where for future convenience we have specialized to the scattering scenario.  In \eqref{BBnumden}, each term appears to correspond to the expectation value of the number of strings produced in the corresponding mode.  For instance, in the scattering of two $p$-branes  the number density after infinite time of the lightest stretched strings (the modes that would be massless for coincident branes, of which there are 32)  is $32  \(\frac{1}{2\pi} \)^p \chi^{p/2} \exp\( \frac{- b^2}{\pi\chi} \)$.

In principle, one could compute  $P_{vac}$ and the string number density directly using open string field theory.  As we will see in \secref{massimo}, at least for the lowest mode the string field theory result agrees with \eqref{BBnumden}.

\subsection{Enhanced production rate} \label{explain chi}
In this section we explore the origin of the factor of $\pi \chi = \(\tanh^{-1}\(\pi e_0 E\) + \tanh^{-1}\(\pi e_\pi E\) \)$
(rather than $(e_0+e_\pi) E = eE$)
that appears in the exponentials in \eqref{BBnumden}:
\be \label{dif}
	\Im[\mathcal{A}_{\text{annulus}}] \propto \exp\(\frac{-r \pi m_0^2}{\chi}\)   \quad \text{versus} \quad   \Im[\mathcal{A}_{\text{Schwinger}}] \propto \exp\(\frac{-r \pi m_0^2}{eE}\) \, .
\ee

We begin by re-deriving the annulus result in the electric case using Euclidean instanton methods.   A similar analysis can be found in \cite{Schubert:2010tx}.
Before considering strings, we review a simple instanton derivation of the Schwinger rate for charged particle production.  
The action for a relativistic charged particle is 
\be
S = \int d\tau \left\{ -\frac{1}{2 \eta} \partial_\tau X^\mu \partial_\tau X_\mu +  \frac{1}{2 }m^2 \eta + e A_\mu \partial_\tau X^\mu \right\}.
\ee
In Euclidean signature $\tau \rightarrow i\tau_E$, $X^0 \rightarrow i X^d$, $A_0 \rightarrow A_d$ and $A_j \rightarrow iA_j$. The action becomes:
\be
S_E = \int d\tau_E \left\{ \frac{1}{2 \eta} \delta_{i j}\partial_{\tau_E} X^i \partial_{\tau_E} X^j  +  \frac{1}{2 }m^2 \eta  + e A_i \partial_{\tau_E}X^i \right\}.
\ee
The equations of motion are:
\be
\begin{split}
m\partial_{\tau_E}^2 X_{i}  -e F_{i j}\partial_{\tau_E}X^j = 0 \\
\eta = m^{-1} \sqrt{\partial_{\tau_E} X^i \partial_{\tau_E} X_i}
\end{split}
\ee
For a constant electric field in the $X^1$ direction, the non-trivial equations are:
\be
\begin{split}
\partial_{\tau_E}^2 X^d =R^{-1} \partial_{\tau_E}X^1 \\
\partial_{\tau_E}^2 X^1 = - R^{-1} \partial_{\tau_E}X^d,
\end{split}
\ee
where $R^{-1} \equiv e E \eta$.  The solution is a circle in the $X^d-X^1$ plane:
\be
\begin{split}
X^d =R \sin (  \tau_E /R)\\
X^1 = R \cos (\tau_E /R) \\
\eta = m^{-1}.
\end{split}
\ee
Plugging this back into the action gives 
\be \label{SEpp}
	S_E = 2 \pi R m  - e E \pi R^2 = \pi { m^2 \over e E},
\ee
which reproduces the leading exponential in Schwinger's  result \eqref{schwing}.  The term $2 \pi R m$ in \eqref{SEpp} is the mass times the length of the worldline, while $e E \pi R^2$ is the field times the charge times the area enclosed by the worldline of the charge.

The solution can be analytically continued to describe a pair of particles that  are at rest at $t=0$ and separated by a distance $2 R$ in the direction of the field, and then undergo constant proper acceleration.  The separation distance $R=m/eE$ is a consequence of conservation of energy, since the electrostatic energy of the charged pair is $-eE \cdot 2R = -2m$.

Now consider an open string with net charge $e = e_0 + e_\pi \neq 0$.  For our purposes it is convenient to consider open strings that stretch between two separated branes a distance $b$ apart.  With zero electric field the energy of such a classical, non-vibrating string is simply $m=b/\pi$ (in units where the string tension is $T_0 = 1/\pi$).  Hence in a  non-zero electric field, Schwinger's result leads  one to expect the rate of string pair production to scale as $e^{-b^2/(\pi e E)}$, rather than $e^{-b^2/\pi\chi}$ as in \eqref{BBnumden}.

However, as we will see just below, from the Euclidian point of view the rate for producing a string pair is enhanced due to the fact that the bulk of the string can re-arrange itself.  Instead of all being concentrated on the circle of radius $R$ where the charged endpoint is, the bulk of the string ``dangles" down to smaller radius where its action cost is lower.

From eq.~\eqref{eq:electric action}, the Euclidean action for a charged string in conformal gauge is
\begin{multline}
	S_E =  \int d \tau_E \int_0^\pi d \sig \bigg\{ \frac{1}{2\pi} \left[ \dot{X^i}\dot{X_i} + X'^{i} X'_i\right] \\
	- \frac{E}{2} \left[ e_0 \del_D(\sig) + e_\pi \del_D(\sig -\pi) \right] ( \dot{X^d} X^1 - \dot{X^1} X^d) \bigg\} \, ,
\end{multline}
where we choose the gauge $A^\mu = - \frac{1}{2} F^\mu_{\phantom{\mu}\nu} X^\nu$, with the electric field in the $X^1$-direction.

For a string stretching between branes separated by a distance $b$ in the $X^2$ direction, the solution to the equations of motion that follow from this action is (see \appref{app:charged}):
\be \label{estringsol}
	X^d = R(\sig) \sin \chi \tau_E \, ,  
	X^1 = R(\sig) \cos \chi \tau_E \, ,  
	X^2 = \frac{b}{\pi} \sig \, ,
\ee
with the other coordinates constant.  Here $   R(\sig) = {b \over \pi \chi} \cosh(\chi_0 - \chi \sigma)$, where $\chi_0 = \tanh^{-1}(e_0 E \pi)$.

The Euclidean action on this solution is
\be \label{SE}
	S_E = \int_0^\pi d \sig \, 2 R(\sig) \sqrt{R^{\prime 2}(\sig) + b^2/\pi^2} - E \pi (e_0 R(0)^2 + e_\pi R(\pi)^2) =  \frac{b^2}{\pi\chi},
\ee
which correctly reproduces the exponent of the annulus diagram.  The first term is the area of the worldsheet times the tension, while the second is the field times the charge times the area enclosed by the worldlines of the charged ends of the string.  

\begin{figure}
\begin{center} 
 \includegraphics[width=.3\textwidth]{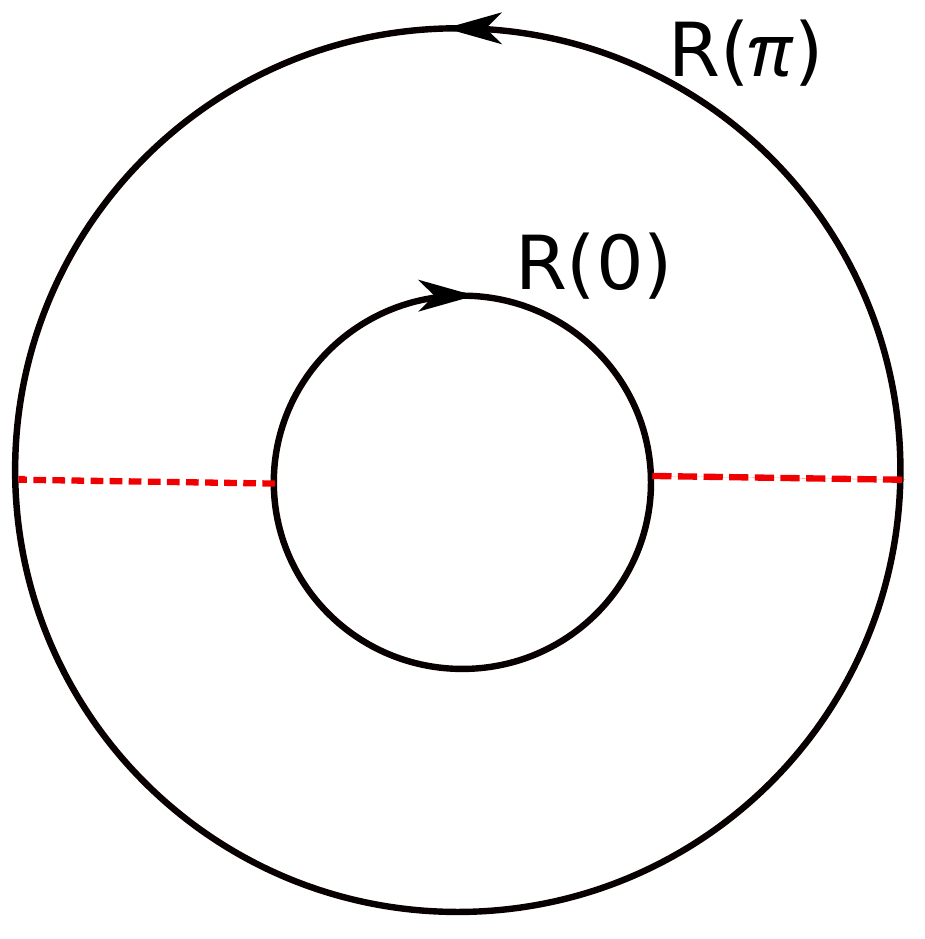}
\end{center}
\caption{\label{flatann}  String instanton computing the rate for charged string pair production in a brane worldvolume electric field.  Cutting the diagram along the red dashed lines reveals the (oppositely oriented) string pair.}
\end{figure}

This is closely analogous to the case of the charged point particle.  For simplicity consider a string with one neutral end ($e_0=0$, for instance).  Then the solution \eqref{estringsol} is an annulus in the $X^d-X^1$ plane, with the worldline of the charged end a circle at the outer radius $R(\pi) = b \cosh(\pi \chi)/\pi \chi$ and the neutral end at the inner radius $R(0) = b/\pi \chi$ (\figref{flatann}).  Applying the formula $R = m/eE$, the radius of the charged end would correspond to a mass $e E R =  b \sinh(\pi \chi)/\pi^2 \chi$.  
In fact this \emph{is} the total mass of the (bulk of) the string, as can be seen from \eqref{eq:charge}.\footnote{That this is \emph{larger} than $b/\pi$ can be understood from the fact that the string curves, rather than stretching straight between the branes.  The force from the electric field requires the charged end of the string to connect to its brane at an angle that depends on the charge times the field, so the string cannot be straight.}  Hence, the contribution to the action from the charged end is identical to that of a charged particle with charge $e$ moving at the radius one would  expect if it had mass $b \sinh(\pi \chi)/\pi^2 \chi$.

However, the crucial difference is that the contribution to the action from the bulk of the string is modified relative to the particle case.
 Each infinitesimal segment of the string with mass $\delta m$ contributes to the action like a particle with that mass: namely it adds $\delta S = 2 \pi R(\sigma) \delta m$, where $R(\sigma)$ is the radius of the circle described by that piece of string.  Because $R(\sigma) \leq R(\pi)$, the bulk of the string contributes less to the action than the charged end does, enhancing the production rate and reproducing the exponent \eqref{dif}.  The moral is that strings are easier to produce than particles, because they have internal degrees of freedom and can arrange themselves to lower the action cost for pair production.

The analytic continuation of \eqref{estringsol} $\tau_E \to -i \tau,  X^d \to -i X^0$ describes a pair of oppositely charged strings at rest at $X^0 = \tau = 0$ that then accelerate in opposite directions.  Using the results of \appref{electric string energy}, one can  check that both the nucleation and subsequent motion of the pair conserves energy.

\subsection{Open string field theory} \label{massimo}

The equations of motion for charged string modes in background fields can be found in \cite{Porrati:2010hm}. 
The physical state conditions for the string state $|\phi\ra$ can be expressed  in terms of the Virasoro generators as
\begin{eqnarray} \label{phys_cond}
	(L_j-\delta_{0j})|\phi\ra &=& 0.
\end{eqnarray}

Focusing on the tachyon of the bosonic string as an example, the only non-trivial equation is $j=0$ in \eqref{phys_cond}, since the rest are trivially satisfied.

Expanded in terms of creation and annihilation operators, $L_0$ is
\be \label{eq:L0}
	L_0 = -\frac{1}{2}\mathcal{D}^2 + \frac{1}{4}\rm{Tr}G^2 + \sum_{m=1}^{\infty}(m+iG)_{\mu\nu}a_m^{\dagger\mu}a_m^{\nu}   ,
\ee
where
\begin{eqnarray}
	G &=& \frac{1}{\pi}\[\tanh^{-1}(\pi e_0 F)+\tanh^{-1}(\pi e_{\pi} F)\]  \\
	\mathcal{D}^{\mu} &=& \sqrt{\frac{G}{eF}}^{\mu}_{\phantom{\mu}\nu}D^{\nu} \\
	F_{\mu\nu} &=& \de_{\mu}A_{\nu}-\de_{\nu}A_{\mu} ,
\end{eqnarray}
and $e = |e_{0}+e_{\pi}|$.  For the tachyon, the last term in \ref{eq:L0} vanishes.

Consider an electric field $E$ in the $X^1$ direction, and choose the gauge $A_1=-Et$. With $\chi = \frac{1}{\pi} [\tanh^{-1}(\pi e_{\pi}E)+\tanh^{-1}(\pi e_0E)]$ as usual, 
\be
	G = \chi \( \begin{array}{cc}
										0 & -1 \\
										-1 & 0 \end{array} \)    ,
\ee
in the $(X^0,X^1)$ subspace, and is zero elsewhere.

Then  \eqref{phys_cond} gives
\be \label{tachyon_eq}
	\left\{-\de_0^2 - 4ieE X^0 \de_1+\nabla^2 -(e E X^0)^2+ \frac{e E}{\chi}\[-\chi^2+2 \] \right\} |T\rangle = 0.
\ee
The ``2" in \eqref{tachyon_eq} is the tachyon mass in bosonic string theory in our units, but here it is multiplied by $eE/\chi$, evidently due to the effect of the electric field.  
The $\chi^2$ arises from $\rm{Tr}G^2$ in \eqref{phys_cond}, and cancels for the superstring due to worldsheet supersymmetry. As usual in the  calculation of the (electric) Schwinger effect, the term involving $\de_1$ can be removed by a $k_1$-dependent time shift once one goes to momentum space; after integrating over momenta this gives rise to the  $T e E = T  | E(e_0 + e_\pi)|$ prefactor in \eqref{eq:annulus}.

If we consider two branes separated by an impact parameter $b$ in the $X^2$ direction, $\mathcal{D}^2$ is shifted:
\be \label{tachyon_eqb}
	\left\{-\de_0^2 - 4ieE X^0 \de_1+\nabla^2 -(e E X^0)^2+ \frac{e E}{\chi}\[-\chi^2+2 - {b^2 \over \pi^2} \] \right\} |T\rangle = 0 \, .
\ee
Referring to \eqref{tmass} and \eqref{VacP}, \eqref{tachyon_eq} correctly reproduces the exponent in the annulus amplitude corresponding to the bosonic string tachyon, which is $e^{2 \pi /\chi - \chi \pi }$ \cite{Bachas:1992bh}, and \eqref{tachyon_eqb} adds $e^{-b^2/\pi  \chi}$ as expected.

  Note that \eqref{tachyon_eq} differs from the proposal of \cite{McAllister:2004gd}.  In particular, the energy of the string at late times goes as $e E X^0$, or $v X^0/\pi$ for brane scattering.   In the latter case, at late times the string is very long and straight and the motion of its endpoints is almost parallel to its extent, even for $b \neq 0$ (\appref{app:scatwork}).  Since the tension and mass density of a relativistic string depend only on its transverse velocity, the energy of such a string should be given by its length at leading order.  In the T-dual electric frame the string has a non-zero net charge, and its endpoints undergo constant proper acceleration.  Therefore the work done by the field on the string is $e E \Delta X^1 \approx e E X^0$, in agreement with \eqref{tachyon_eq} (see \appref{electric string energy}).


\section{Brane scattering} \label{bscat}

 Once pairs of open strings are produced they create a force that binds the branes together.
 This is the stringy version of the moduli trapping mechanism of \cite{Kofman:2004yc}, and after the scattering it eventually brings the branes to a stop, or potentially into some sort of orbit in the case $b \neq 0$ (in the electric frame, this deceleration corresponds to the decrease of the electric field as a result of charged pair production).  
 
 The scattering dynamics  is quite complex in general, so in this section we will focus on a simple proxy -- the {\bf stopping distance} $l_{*}$ as a function of velocity.  That is, the distance the branes move apart in the center of mass frame after a $b=0$ collision before the energy in stretched strings equals the brane kinetic energy.  (In the electric frame this  corresponds to the time it takes to discharge of the  field.)  As we will see, a surprising feature of brane scattering is that at least for sufficiently high Lorentz factor, the stopping distance \emph{decreases} as a function of \emph{increasing} velocity (as previously noted in \cite{McAllister:2004gd}).

One must be cautious in pushing our method of analysis too far.  The annulus result allows us to infer the number density of strings a long time after the scattering, in the approximation of constant brane velocity.  This means that significant changes in the brane velocity as a result of string production will invalidate the analysis.  Furthermore, string production is a quantum process, and not all pairs of strings are produced at the moment of closest approach (or indeed, at any definite time).  Hence one cannot be sure precisely when or at what brane separation to begin including the force due to produced strings.  Lastly, the energy in strings with moving end points is subtle, as we have seen in the previous section in the electric case.

Fortunately, all of these issues can be dealt with as long as we remain in a certain parametric regime.  Strings are produced when their masses are changing most rapidly, namely when the branes are relatively close together.  Furthermore, as we establish carefully in \appref{app:scatwork} the energy in a very long string with endpoints moving in the direction nearly parallel to its length is at leading order simply equal to its length (divided by $\pi$ in our units).  This is the case for the stretched strings at late times, and so  the force all strings exert at late times is simply $1/\pi$ -- independent of the brane velocity, and independent of the string mode.\footnote{Our analysis differs from \cite{McAllister:2004gd} on this point.}   In other words any uncertainty in the string force arises when the branes are relatively close together, and in this regime we also do not know precisely when the strings are produced.

\subsection{Stopping distance} \label{sl}
 We can trust the number densities  \eqref{BBnumden} well after the collision when the branes are far apart.
   At sufficiently large separations, $l$, the energy per string is close to $l/\pi$.  Therefore, in this limit the energy density in strings produced during brane-brane scattering is approximately
\be \label{rhos}
	\rho_s =   \frac{l}{ \pi}  \(\frac{1}{2\pi} \)^p \chi^{p/2}  
	  \left\{ 32 + 512 \exp\(-\frac{2\pi}{\chi}\) + 4608\exp\(-\frac{4\pi}{\chi}\) + \cdots \right\} 
\ee 
where we have used equation \eqref{BBnumden}.   The corresponding formula for brane-antibrane scattering  uses the second line of \eqref{BBnumden} in the obvious way.

\paragraph{Non- and moderately relativistic velocities:}

In the non-relativistic regime, open string production is exponentially suppressed for all massive modes.  However if $b < 1$ the lightest modes (massless and tachyonic) are still produced copiously -- and at sufficiently low velocities in brane-antibrane scattering the tachyon can condense and the branes annihilate into closed strings (c.f.~\secref{ann}).  

\begin{figure}
\begin{center} 
 \includegraphics[width=1\textwidth]{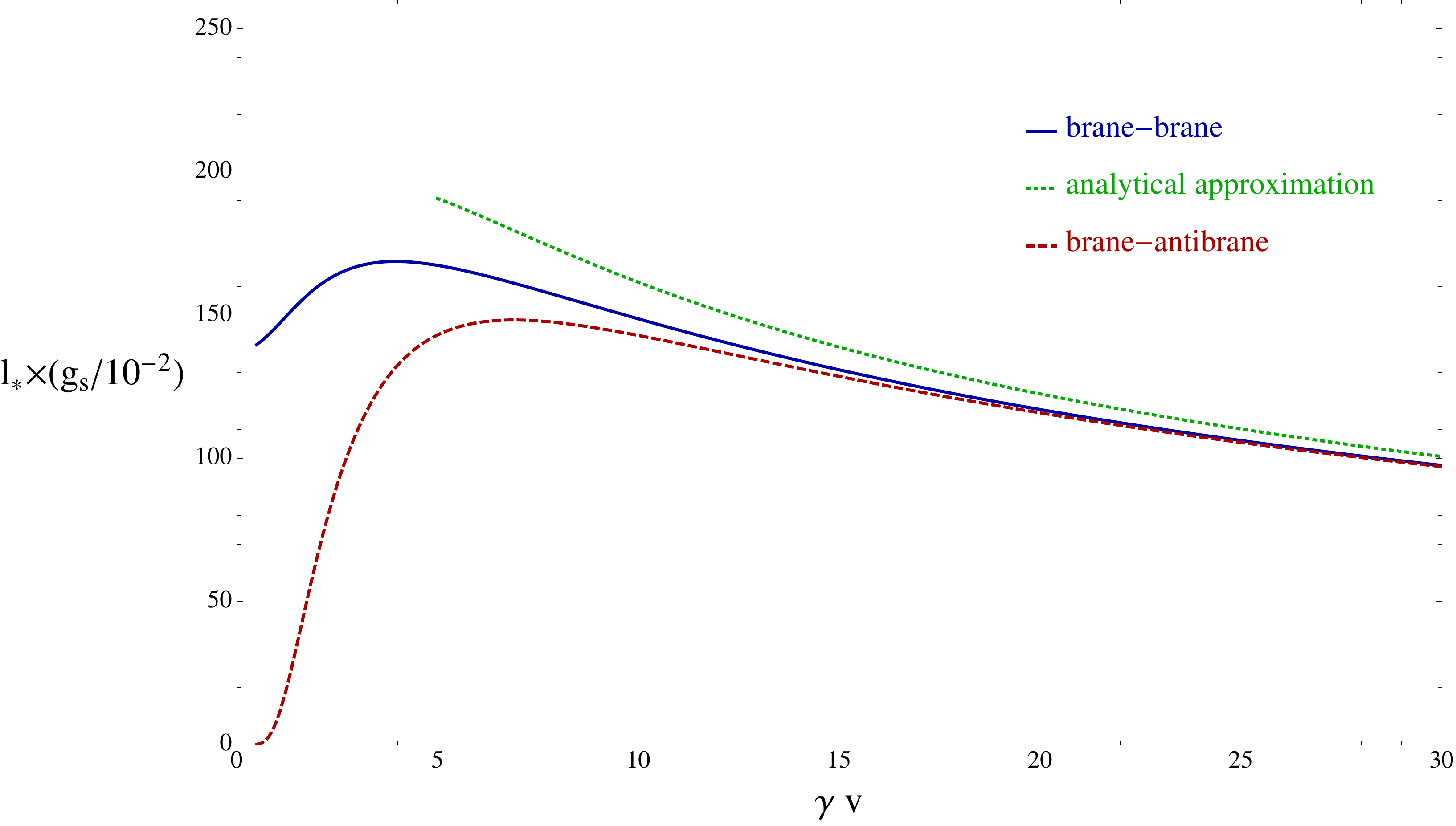}
\end{center}
\caption{\label{stopdist}The stopping distance $l_*$ in units $\alpha'=1/2$ in the center of mass frame for the scattering of two 4-branes, or a 4-brane  anti-4-brane pair, as a function of the Lorentz factor  $ \gamma$ times the speed $v$ of either brane.  The lower two curves are computed by equating \eqref{rhos} to the initial brane kinetic energy and should be accurate for all $\gamma v$ at sufficiently small $g_{s}$; the top curve is \eqref{stopdistmk}, accurate for large $\gamma$.  The stopping distance scales linearly with the D-brane tension $\sim g_{s}^{-1}$.}
\end{figure}

\begin{figure}
\begin{center} 
 \includegraphics[width=.92\textwidth]{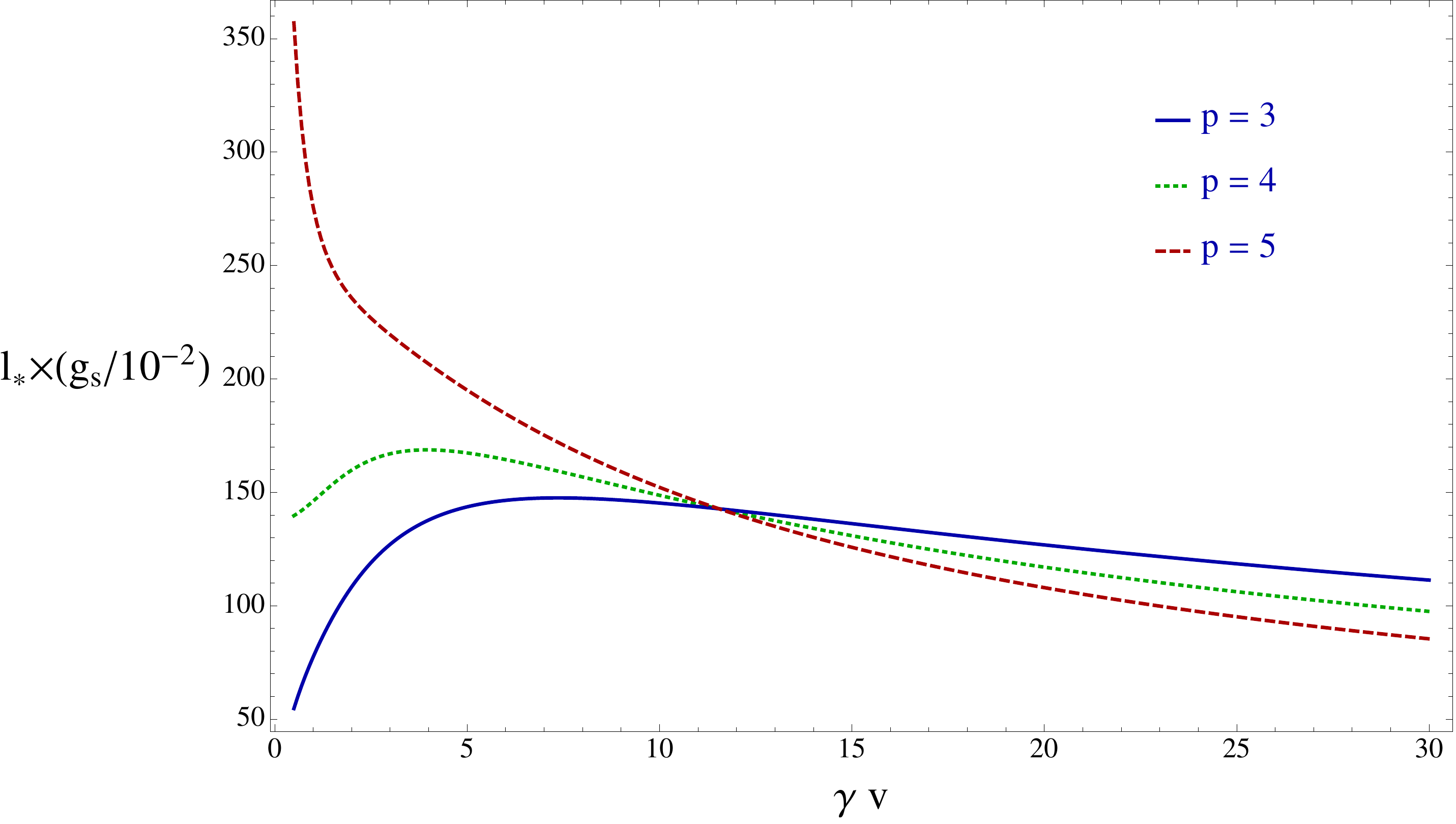}
\end{center}
\caption{\label{stopdistp}The stopping distance $l_*$ in units $\alpha'=1/2$ in the center of mass frame for the scattering of two $p$-branes, for $p=3,4,5$, as a function of the Lorentz factor $\gamma$ times the speed $v$ of either brane, computed by equating \eqref{rhos} to the initial brane kinetic energy.  The stopping distance scales linearly with the D-brane tension $\sim g_{s}^{-1}$.}
\end{figure}

For $\gamma \sim \O(1)$ no simple analytic approximation to \eqref{rhos} is available, but only the first few terms in the sum in \eqref{rhos} are relevant so there is no difficulty in finding the stopping distance numerically.  The results are plotted in \figref{stopdist} and \figref{stopdistp}.  
For brane-brane scattering with $p>4$ and for all $p$ at $\gamma v \simgeq 10$, the stopping distance \emph{decreases}  with increasing velocity.  This counter-intuitive behavior results from the fact that at higher velocities, more and more massive modes are produced (since the suppression $e^{-m^{2}/\chi}$ becomes less relevant at larger $\chi$), and the rapidly growing degeneracies mean that this effect is so strong it more than compensates for the additional brane momentum.

 At sufficiently small values of $\gamma v$ in brane-brane scattering only the massless modes are relevant,  so from \eqref{rhos}, one has $\rho_{s} \sim \chi^{p/2} \sim v^{p/2}$.  This combined with the fact that  kinetic energy scales as $v^{2}$ for small $v$ explains the $p$-dependent behavior plotted in  \figref{stopdistp}.  The same qualitative behavior occurs for brane-antibrane scattering for any $p$.  In that case at low velocities the tachyonic contribution to the string energy  density $\sim \chi^{p/2} e^{+\pi /\chi}$  dominates, which is a rapidly decreasing function of $v$ in the small $v$ regime.

\paragraph{The ultra-relativistic limit:}   
For either branes or a brane-antibrane pair, the density of states $\nu(j)$ at  large $j$ can be approximated by
\be \label{MultipApprox}
	\nu(j) \approx j^{-11/4}e^{2 \pi\sqrt{j}} .
\ee
Using this expression,  \eqref{rhos} (and its equivalent for brane-antibrane) can be approximated by
\be
	\rho_s  \approx  \frac{l}{ \pi}  \(\frac{1}{2\pi} \)^p \chi^{p/2}  \int^{\infty}d j \,  j^{- 11/4} \exp\(2 \pi\sqrt{j} - \frac{\pi j}{\chi}\).
\ee
The exponential increase in the multiplicity of states combines with the exponential suppression at large $j$ so that the integral is peaked at  $j_{\rm peak} \approx \chi^2$.  
For large $\chi$ the integral can be approximated by:
\be
	\int^{\infty}dj \,  j^{- 11/4} \exp\(2 \pi\sqrt{j} - \frac{ \pi j}{\chi}\) \approx \frac{ e^{\pi \chi} }{ \chi^4} \, .
\ee
Using this approximation to rewrite \eqref{rhos} we have
\be \label{rhosapprox}
	\rho_s  \approx {l \over \pi} \(\frac{1}{ 2\pi}\)^p \chi^{p/2-4} e^{\pi \chi} \,.
\ee
The kinetic energy density of the brane pair is $2 (\gamma-1)$ times the brane tension $\tau_p$
\be
	\rho_{D_p} =2 (\gamma-1) \times \tau_p = 2 (\gamma-1) \times {2^{(p+1)/2} \over g_s(2\pi)^p} \approx \frac{2^{(p+1)/2}}{g_s(2\pi)^p}e^{\pi\chi/2} \, .
\ee
Setting $\rho_{D_p} = \rho_{s}(l_*)$ one finds the stopping length 
\be \label{stopdistmk}
	l_* \approx {   2^{p+1 \over 2}\pi} {\chi^{4-p/2} e^{-\pi \chi/2} \over g_s}  \approx  2^{p-1 \over 2} \pi \frac{   \chi^{4-p/2}}{ \gamma g_s} \, .
\ee
For $\chi \gg 1$ (where the approximations used to derive \eqref{stopdistmk} are valid) this is a \emph{decreasing} function of $\chi$.  In other words, the branes exhibit the counterintuitive behavior that the stopping distance decreases with increasing initial velocity \cite{McAllister:2004gd}.  

\subsection{Brane-antibrane annihilation} \label{ann}

Despite the fact that there is a tachyon in the open string spectrum when the brane and antibrane are within a string length,  brane-antibrane scattering at small impact parameter will not necessarily lead to annihilation.  The  reason is that at high velocity the branes spend very little time within a string length of one another, so that the tachyon has limited time to condense \cite{Kleban:2011cs,D'Amico:2012ji}.  Nevertheless, the   phenomenon described in \secref{sl} will strongly bind the brane-antibrane pair at sufficiently relativistic velocities, stopping them rapidly and presumably then allowing the tachyon to condense.  By contrast at low velocities, the tachyon has ample time to condense.  For this reason we will find a range of velocities that is bounded from both below and above where the brane-antibrane pair can pass through or near each other without annihilating.

The decay of the tachyon in a brane-antibrane system is a nonperturbative process that is not well-understood.  However for our purposes, the relevant physics can be captured by an effective action, and as we will see the details of the action are  not very important for what we want to establish.  In the remainder of this section we will be concerned with the non-relativistic limit, $v \approx \pi \chi, \gamma \approx 1$.

One concrete model is the proposal of~\cite{Minahan:2000tf}, valid for a static brane-antibrane system:
\be \label{minac}
	S = - 8 \tau_p \int d t d^p x \[ \frac{1}{2} e^{- 2 |y|^2} |\de_\mu y |^2 + {1 \over 4} e^{- 2 |y|^2} \] .
\ee
Here $y$ is the (dimensionless) complex tachyon field and $\tau_p$ is the D-brane tension. The potential has a maximum at $|y|=0$ and a minimum at $|y| = \infty$, which actually corresponds to a finite distance $\sim g_{s}^{-1/2}$ in field space for the canonically normalized field.  The energy difference between the maximum and the minimum is twice the tension of the D-brane.

If we consider  scattering branes with non-relativistic velocity $v$ and impact parameter $b=0$, a simple proposal is to modify the potential \eqref{minac} as follows:
\be \label{minmod}
	S = - 8 \tau_p \int d t d^p x \[ \frac{1}{2} e^{- 2 |y|^2} |\de_\mu y |^2 + e^{- 2 |y|^2}  \left\{ {1 \over 4} +  {1 \over 2} \({ v t \over \pi } \)^2  |y|^2 \right\} \] \, .
\ee
For $t=0$ this coincides with $\eqref{minac}$, but the second derivative at $y=0$ is modified to coincide with the time-dependent mass of the tachyon.  The potential for the canonically normalized field still has global minima at a finite distance $g_s^{-1/2}$, with an energy difference $2 \tau_p$ from $y=0$, but now in addition has maxima at sufficiently early and late times, with a height that grows with $|t|$.  These represent the tunneling barriers for the tachyon to condense when the branes are separated by more than a string length.

If we start at $t \to - \infty$, the (vacuum state) wavefunction will be concentrated at $y = 0$.
Expanded around this point, the  action is
\be
\label{eq:free}
	S_2 = - \int d t d^p x \( |\de_\mu \phi|^2 +  2 \tau_p  + \left( -1 +\({ v t \over \pi } \)^2  \right)  |\phi|^2 + {\mathcal O}(g_{s} |\phi|^{4})\)  ,
\ee
where $\phi \equiv (4 \tau_p)^{1/2} y$ is  canonically normalized up to non-linear corrections.  As promised, the mass has the correct time dependence.  The only significant input from \eqref{minac} is the generic feature that the first non-linear term is of order $g_{s} |\phi|^{4}$ (and so non-linearities become important when $|\phi| \sim 1/\sqrt{g_s}$).
It is in this sense that the details of the effective action for the tachyon are not important for our analysis -- we will only make use of the quadratic action and measure the variance of the wavefunction against the scale of non-linearities $1/g_{s}$.

As time goes on, the field becomes lighter and lighter until, around $t = 0$, it becomes tachyonic and can decay.  Here by ``decay'' we mean that the field takes a value $|\phi|^{2} \sim 1/g_{s}$ where non-linearities become important.  
We would like to  calculate the decay probability, per unit volume, as a function of the velocity of the branes.

The quadratic theory  \eqref{eq:free} has the exact solution:
\be
	\hat{\phi}(x) = \int d_p k \[ e^{i \k \cdot \x} u_k(t) \hat{a}_{\k} +  e^{-i \k \cdot \x} u^\ast_k(t) \hat{b}^\dag_{\k}\] 
\ee
where we take into account that the field is complex.
The mode functions satisfy
\be
	\ddot{u}_k + \left[ k^2 - 1+  \( v t /\pi\)^2 \right] u_k = 0 \, .
\ee
The properly normalized solution which corresponds to the \emph{in} vacuum at past infinity is
\be
	u_k(t) =  (2 v/\pi)^{-1/4} e^{\frac{\pi^2}{8  v } (1 - k^2)} D_\lam\( (i-1)  t\sqrt{ v/\pi} \) \,
\ee
where $D_p(z)$ is a parabolic cylinder function and the index is $\lam = -\frac12 + i \pi \frac{(k^2 - 1)}{2  v}$. 

The two-point function in Fourier space is simply
\be
	\avg{\hat\phi_{\k} \hat\phi^\dag_{\k'}} = \del_D(\k + \k') |u_k(t)|^2 \, .
	\ee
As long as this is small compared to $1/g_{s}$ we can trust the quadratic approximation, and the wavefunction for the mode will be Gaussian with this variance. However, we are not interested in the $k$-modes themselves.  Rather, we want the probability that in some region of spatial volume $R^p$ the field reaches the non-linear regime and decays.
To find this, we smear the field and compute the 2-point function of the smeared field:
\be
	\phi_R(x) = \int d^p y W_R(|x-y|) \phi(y) = \int d^p k \, \tilde{W}(k R) \phi_{\k},
\ee
where $W_R(x)$ is a filter function such as a Gaussian or a top-hat.
The variance of $\phi_R$ is 
\be
	\avg{|\phi_R|^2} \equiv \sig_R^2 = \int d^p k \, \tilde{W}^2(k R) |u_k(t)|^2
\ee
which is independent of $x$ by translation invariance.

For a generic filter function, we can estimate this integral by replacing $k \to 1/R$ and dividing by the volume factor:
\be \label{varR}
	\sig_R^2  \approx \frac{1}{R^p} |u_{1/R}|^2 \, .
\ee
The probability that a region of volume $R^p$ decays is the probability that the field is above a critical value, which we take to be the position of the maximum of the  barrier in \eqref{minmod}, which is $\phi \approx 1/\sqrt{g_s}$.

Eq.~\eqref{varR} is easy to evaluate numerically. 
For large negative times it is small, as expected because there is no particle production then.
At $t \sim 0$ it begins to increase, reaches a maximum at $t \sim \pi/v$, and then oscillates with a decreasing envelope at large positive time.  The maximum variance  is well approximated (for $R \gg 1$) by
$$
\sig^{2}_{R, max} \approx R^{-p} \exp(\pi^2/v)
$$
where the approximation is good at small $v$.

Therefore, conservatively using the maximum value of $\sigma(t)$, the probability that a region of size $R$ decays is 
$$
P_{R} \approx {\rm erfc}\[{1 \over \sigma \sqrt{2 g_s}}\] \approx \exp\[-{ 1  \over  2 g_s \sig^{2}_{R, max}} \] \approx  \exp\[{-R^{p} e^{-\pi^2/v} \over 2 g_{s}}\].
$$
The main feature of this formula is the dependence on $g_{s}$, which shows that for fixed $R,v$ the probability goes exponentially to zero as $g_{s} \to 0$.

For the relativistic regime, so long as the stopping length $l_*$ computed in the previous section exceeds the string length, the tachyon will not condense at least on the the first pass (although the branes may be pulled back together and subsequently annihilate).  Therefore we have established what we set out to show -- that the probability for annihilation can be made small in the limit $g_{s} \to 0$.  One sees that the annihilation probability is small so long as $v \simgeq \pi^2 / | \ln g_{s} |$ and $\gamma \simleq 1/g_s$ (c.f. \eqref{stopdistmk}), or simply
$$
{\pi \over |\ln g_s|} \simleq \chi \simleq {2 \over \pi} | \ln g_s |.
$$

\section{Annihilation and reheating in unwinding inflation} \label{uwi}

One motivation for this work is unwinding inflation, where slow-roll inflation occurs due to the gradual unwinding of  a higher-form electric flux.  Here we will only comment briefly on brane-antibrane annihilation and its relevance to reheating in this model.  For brevity we will not review the model now; the reader can refer to \cite{D'Amico:2012sz} for a brief, self-contained description, and to \cite{D'Amico:2012ji} for more detail, including some comments on reheating and tachyon condensation.

Unwinding inflation requires that a spherical brane repeatedly self-intersect without immediately annihilating as it expands around a compact direction(s).  Locally after a few efolds of inflation, the brane's radius is large and hence the self-intersections are well approximated by a planar brane-antibrane collision.  Since the de Sitter radius is much larger than the string length, flat space should be a good approximation within a few string times of the collision.  Inflation ends and reheating occurs when the brane slows down enough (which happens naturally as the flux is reduced) and self-annihilates.

There is however a crucial difference relative to a flat space collision, due to the presence of a background flux.  If the brane annihilates in a region where some flux remains, this region will have a larger energy density than regions where the unwinding process continues and the flux is completely discharged, or reduced to a lower level.  Such high-energy regions will collapse into small black holes unless they are either larger than the Hubble length or so dense that they percolate.  

This leaves two possibilities.  One is that unwinding inflation will end when all the flux has been discharged, with a few rare regions where an ``undershoot" or ``overshoot" led to annihilation with some residual flux.  In this case, these regions will look like primordial black holes produced during reheating, and will evaporate in much less than the life of the universe.  

The other possibility is that the branes will annihilate before all the flux has discharged (or even after -- ``momentum" can carry the discharge process past zero in some cases).  In that scenario, one expects the regions with the smallest amount of flux remaining to expand, since their energy density is lowest.   To see what can happen, suppose $p_{N+1} \approx 1$, $p_N \approx e^{-180}$, and all other $p_i=0$, where  $p_N$ is the probability for an inflationary Hubble region to annihilate and reheat with $N$ units of flux remaining.  Then in the $\sim e^{3 \times 60}$ Hubble volumes at the end of inflation that are visible today, nearly all will have $N+1$ units of flux, but it is likely that a single region will have $N$ units of flux instead.  In that case the region with $N$ units of flux will expand rapidly, much like in old inflation, and this will almost certainly lead to a cosmology that is inconsistent with observation.  On the other hand, if $p_N \ll e^{-180}$ there are unlikely to be any such regions.

Now consider instead the case where $p_{N+1} \sim 1, p_{N} \sim e^{-135} = \(e^{15}\times e^{-60}\)^3 \gg e^{-180}$.  In this case, the characteristic separation between the rare regions with $N$ units of flux will be $\sim e^{-15}$ times the horizon scale today, in other words a few thousands of parsecs.  When these bubbles collide they will produce large primordial density fluctuations on that scale.   However, because the CMB can only probe roughly 8 efolds of scale starting from the dipole, and other direct probes of primordial perturbations extend this by only a few more efolds, such peaks in the primordial spectrum are very poorly constrained.

Hence, it seems that there would be an (easily) observable signature only if $p_0$ is very small and there exists an $N$ such that
$$
e^{-135} \simleq p_N \simleq e^{-180},
$$
(where the numbers ``135" and ``180" are uncertain at $\O(1)$).  This interesting conclusion will be investigated further in future work.

\section{Conclusions} \label{conc}
There are many open questions remaining to be investigated.  We list a few below.

From our point of view (again motivated by unwinding inflation), one of the most interesting is the question of what happens to a spherical D-brane in flat spacetime that initially has a large radius.  Such a brane will collapse to a point as a result of its tension.  An F-string at $g_s=0$ would simply re-expand (with reversed orientation), and continue to oscillate indefinitely, reversing orientation each time.  If the brane  passes through itself without annihilating in a similar way, it can ``unwind'' the field it is electrically coupled to (c.f. Appendix A of \cite{D'Amico:2012ji}).  But for a D-brane the situation is more complicated than for an F-string even (or especially) at small $g_s$.  Will the brane self-annihilate on the first pass, or will it simply produce some strings that have a small effect on its motion, as in the planar case we have analyzed here?  

In the relativistic limit, the energy in a classical stretched string at the moment of closest approach of the branes is $\sim (b/\chi) e^{\pi \chi}$  (c.f. \eqref{stringen}).  This is greater than $b$ because  the transverse velocity of the string increases its effective mass density, and it is also larger than the energy implied by the equations of motion of string field theory  or equivalently by the rate of production of string pairs, which is $\sim b \sqrt{v/\chi}$ (c.f. \eqref{tachyon_eqb}).  While this difference is irrelevant for the dynamics of brane scattering in the limit $g_s \to 0$, it is important if one wants to accurately estimate the stopping length (or rate of discharge of the field) at finite $g_s$.  While our analysis partially clarifies this issue, there is likely more to be learned from investigating it.

Another issue concerns closed string radiation and the real part of the annulus diagram.  The analysis of \cite{Bachlechner:2013fja} shows that the power in closed string Bremsstrahlung grows like a high power of $\gamma$.  Hence at fixed $g_s$, the ultrarelativistic limit of brane scattering will be dominated by closed string radiation rather than open string production.  The interplay between these two and the resulting dynamics remain to be investigated.

More generally, it would be very interesting to study brane scattering at high energies near the black hole formation threshold, or compare it to studies of string brane scattering such as \cite{D'Appollonio:2010ae}.

\section*{Acknowledgements}
It is a pleasure to thank T.~Bachlechner, S.~Dubovsky,  R.~Flauger, B.~Freivogel, V.~Gorbenko, J.~Maldacena, L.~McAllister, M.~Porrati, E.~Silverstein, and G.~Veneziano for discussions.
The work of G. D'A. is supported by NASA through grant NNX10A171G and by NSF through AST-1109432.
The work of MK, RG, and MS is supported in part by the NSF  through grant PHY-1214302 and by the John Templeton Foundation.  The opinions expressed in this publication are those of the authors and do not necessarily reflect the views of the John Templeton Foundation.

\appendix

\section{Production of scalar particles}
\label{app:scalars}

In this appendix, we compute the rate of production for scalar particles with a time-varying mass
\be
	m^2(t) = m_0^2 + A^2t^2 \, .
\ee
Expanding the scalar field in $k$-modes,
\be
	\phi(x) = \int d^{d-1} k e^{i \vec{k}.\vec{x}}\phi_k(t) =  \int d^{d-1} k e^{i \vec{k}.\vec{x}} \left( u_k(t) a_k + u_k^*(t) a_{-k}^\dagger \right),
\ee 
the Klein-Gordon equation reduces to:
\be
	\ddot{u_k} + (m_0^2 + k^2 + A^2t^2)u_k = 0.
\ee
The most general solution for $u_k$ is:
\be
	u_k = C_1 D_{-\nu -1}(z) + C_2 D_{\nu}(i z)
\ee
Where $D_\nu(z)$ is a parabolic cylinder function, $\nu = -1/2 + i m_k^2/(2A)$, $z = (1+i)\sqrt{A}t$, and we have defined $m_k^2 = m_0^2 + k^2$.  We now define two independent sets of mode functions:
\be
\begin{split}
	u_k^{in}  &= \frac{e^{-\pi m_k^2/(8A)}}{(2A)^{1/4}} D_\nu(iz) \\
	u_k^{out}  &= \frac{e^{-\pi m_k^2/(8A)}}{(2A)^{1/4}} D_{-\nu-1}(z), \\
\end{split}
\ee
where $u_k^{in}$ ($u_k^{out}$) is chosen to have positive frequency in the asymptotic past (future), and the constants $C_{1,2}$ are chosen by enforcing canonical commutation relations, $[\phi, \dot{\phi}]=i$.
From this we can find the Bogolubov coefficients
\be \label{Bogo deff}
	a_{\k}^{out} = \alpha_{kk'} a_{\k'}^{in} + \beta_{kk'} a_{-\k '}^{in \dagger}, 
\ee
to be
\be \label{Bogo coeff}
\begin{split}
	\alpha_{kk'} &= \frac{\sqrt{2\pi} \exp\left(\frac{i\pi A-\pi m_k^2}{4A}\right)}{\Gamma(1/2 - i m_k^2/(2 A))} \delta_{kk'} \\[12pt]
		\beta_{kk'} &= \exp\left(\frac{i\pi A-\pi m_k^2}{2A}\right)  \delta_{kk'}.
\end{split}
\ee
One can easily see that
\be \label{nk}
	\langle in | n_k^{out} | in \rangle= \langle in | a_{\k}^{out \dagger}a_{\k}^{out} | in \rangle = |\beta_k|^2 = \exp\left(\frac{-\pi m_k^2}{A}\right),
\ee 
meaning that the in-vacuum contains (on average) $|\beta_k|^2$ particles of the out k-mode.
The total number of particles is:
\be \label{num produced}
	\langle n \rangle = \left( \frac{L}{2\pi} \right)^{d-1}\int d^{d-1}k \langle n_k \rangle =  \left( \frac{L}{2\pi} \right)^{d-1} A^{(d-1)/2} \exp\left(\frac{-\pi m_0^2}{A}\right)
\ee

In \secref{prodrate} we derived the vacuum persistence probability from the number density \eqref{nk}.  We will do so again here by a slightly different method.  To begin, we need the probability of producing $n$ pairs with a wave number $\pm \vec k$:
\be \label{n pair prob}
	P_n(k) = |\langle in | \frac{(a_{\k}^{out \dagger}a_{-\k}^{out\dagger})^n}{n!}|out\rangle |^2.
\ee
To express the in-vacuum in terms of the out-Hilbert space, note that the in-vacuum for the momentum modes is Gaussian, and in a free theory can only evolve into another Gaussian.  The most general  Gaussian is a squeezed coherent state, but a coherent state would violate conservation of momentum.   A careful calculation using \eqref{Bogo deff} and \eqref{Bogo coeff} yields the squeezed state
\be
	|in\rangle = C_0 \exp\left( \int \frac{d^{d-1}q}{(2\pi)^{d-1}} \frac{\beta_q}{2\alpha_q^*} a_{\vec q}^{out \dagger} a_{-\vec q}^{out \dagger} \right) |out\rangle.
\ee
Plugging this in to \eqref{n pair prob} one finds,
\be 
	P_n(k) = |C_{0,k}|^2 \, |\left( \beta_k / \alpha_k^*\right)^n |^2,
\ee
and using,
\be
	\sum_{n=0}^\infty P_n(k) = |C_{0,k}|^2 \frac{1}{1-|\frac{\beta_k}{\alpha_k}|^2} = 1,
\ee
we find
\be \label{P0 scalar}
	|C_{0,k}|^2 = P_0(k) = \frac{1}{1+|\beta_k|^2} =  \frac{1}{1+\langle n_k \rangle} .
\ee

The vacuum persistence probability is the probability that we never produce any particles of any wavenumber, i.e.:
\be \label{Pvac}
	P_{vac} = \prod_{\vec k/{\bf Z}_2} P_0(\vec k) = \exp\left[{1 \over 2} L^{d-1} \int \frac{d^{d-1}k}{(2\pi)^{d-1}} \, \ln(P_0(k)) \right] \, ,
\ee
where the product is over half the momentum space because particles are always produced in pairs with momentum $\pm \vec k$.  Substituting \eqref{P0 scalar} into \eqref{Pvac}, one finds:
\be
\begin{split}
	\ln(P_{vac}) &= -{1 \over 2} \left(\frac{L}{2\pi}\right)^{d-1} \int d^{d-1}k \, \ln(1+\langle n_k \rangle) \\
	&=-{1 \over 2} \left(\frac{L}{2\pi}\right)^{d-1} \frac{2\pi^{(d-1)/2}}{\Gamma\left(\frac{d-1}{2}\right)} \int dk \, k^{d-2} \sum_{n=1}^\infty \frac{(-1)^{n+1}}{n} \exp\left(-n \pi (k^2+m_0^2)/A\right) \\ 
	&= - {1\over 2} \left( \frac{L}{2\pi}\right)^{d-1} A^{(d-1)/2}\sum_{n=1}^\infty \frac{(-1)^{n+1}}{n^{(d+1)/2}}\exp\left( \frac{-n \pi m_0^2}{A} \right), 
\end{split}
\ee
which reproduces \eqref{VacP} as promised.

\section{Charged strings in a constant external electric field} \label{app:charged}

Here we summarize some relevant results from \cite{Porrati:2010hm} and calculate the energy of a charged bosonic string in an electric field.
The action of a charged string coupled to a $U(1)$ field $F_{\mu\nu} = \partial_\mu A_\nu - \partial_\nu A_\mu$ reads:
\be \label{eq:electric action}
	S= \frac{1}{2\pi}\int d\tau d\sig \( \dot{X^\mu} \dot{X_\mu} - X'^\mu X'_\mu \) + \int d\tau d\sigma \( e_0\delta(\sig) + e_\pi \delta(\sig -\pi)\) A_\mu \dot{X^{\mu}} \ .
\ee
The units are chosen such that the string tension is $T = 1/\pi$ ($\alpha' = 1/2$) and $e_{0,\pi}$ are the charges at the two endpoints of the string.
The coupling to the field is a boundary term, therefore the equations of motion are the ones for a free string
\be
	\ddot{X}_\mu - X''_\mu = 0 \, ,
\ee
but with non-trivial boundary conditions
\begin{eqnarray} 
	X'_\mu =& - \pi e_0 F_{\mu \nu} \dot{X}^\nu&    (\sigma = 0) \label{boundc1} \\
	X'_\mu =& \pi e_\pi F_{\mu \nu} \dot{X}^\nu&   (\sigma = \pi) \label{boundc2} \ .
\end{eqnarray}

The full solution of the equations above reads~\cite{Porrati:2010hm}:
\be
\label{eq:solPRS}
	X^\mu(\tau,\sig) = \ x^\mu + \[\alpha_0\(\frac{e^{-G_0}}{2}\cdot \frac{e^{G(\tau+\sig)}-M_+}{G}+\frac{e^{G_0}}{2} \cdot \frac{e^{G(\tau-\sig)}-M_-}{G}\)\]^\mu  + {\rm oscillators}
\ee
where
\be
	G_{0,\pi}=\tanh^{-1}\(\pi e_{0,\pi}F\) \qquad \textrm{and} \qquad G = \frac{1}{\pi}(G_0+G_\pi) 
\ee
and
\be
	M_\pm = \sqrt{\frac{G}{(e_0+e_\pi)F}} \, \sech\(\frac{G_\pi-G_0}{2}\)e^{\pm \pi G/2} \ ,
\ee
and $\alpha_0^\mu = [G/(eF)]^\mu_{\, \nu} p^\nu$.
The expression~\eqref{eq:solPRS} has a smooth limit in the neutral string limit $e_\pi \to - e_0$. For more details, we will refer the reader to~\cite{Porrati:2010hm}.
In addition we  need to impose the constraints, which are those of the free theory:
\be \label{constraint}
	(\dot{X} \pm X')^2 = 0 \, .
\ee

Consider a constant electric field $E$ in the $X^1$ direction and impose Dirichlet boundary conditions in the $X^2$ direction, so that the string is stretched between two D-branes separated by $b$ in the $X^{2}$ direction.
Without loss of generality we focus on a solution in which the brane is at rest at $\tau=0$, which forces $p_1=0$.
The solution for the zero modes is then
\be
\begin{split}
\label{eq:electricsol}
	X^0 &= x^0 + \frac{p^0}{\sqrt{E (e_0 + e_\pi) \chi}} \cosh(\chi_0 - \chi \sigma)\sinh(\chi \tau) \\
	X^1 &= x^1 - \frac{p^0}{E (e_0 + e_\pi)}  + \frac{p^0}{\sqrt{E (e_0 + e_\pi) \chi}}  \cosh(\chi_0 - \chi \sigma) \cosh(\chi \tau) \\
	X^2 &= x^2 + \frac{b}{\pi} \sigma \\
	X^i &= \textrm{constant} \, \, {\rm for} \, \,  i>2,
\end{split}
\ee
where  $\chi_0=\tanh^{-1}(\pi e_0 E)$.
Enforcing the constraint \eqref{constraint} fixes $p^0$ to:
\be
	p^0 = \pm \frac{b}{\pi} \sqrt{\frac{E(e_0 + e_\pi)}{\chi}} \, .
\ee

\subsection{String Energy} \label{electric string energy}
We want to compute the energy associated to a classical string in an electric field. To do so, we calculate the Noether charges associated with translation symmetry in the action \eqref{eq:electric action}, namely we will find the charges associated with the transformations $X^\mu \rightarrow X^\mu +\epsilon^\mu$. The conserved charge associated with translation of $X^0$ is what we will call energy.

The free part of \eqref{eq:electric action} is trivial and gives the expected contribution to the current while the boundary terms add a non-standard contribution, giving
\be
	\delta S =  \int d\tau d\sig \left\{ \frac{1}{\pi} \left( \dot{X^\mu} \dot{\epsilon_\mu} - X'^{\mu}  \epsilon_\mu '  \right)-\( e_0 \delta(\sig) + e_\pi \delta(\sig -\pi)\) F_{\mu \nu} X^\nu \dot{\epsilon^\mu} \right\}\  ,
\ee
where we have used the fact that $F_{\mu \nu}$ is constant to integrate by parts. From this expression we can read off the components of the conserved Noether currents:
\be
	(P^\mu_\tau,  P^\mu_\sig) = \( \frac{1}{\pi}\dot{X^\mu}  - e_0 F^\mu _{\, \, \,  \nu} X^\nu(\sig=0) - e_\pi F^\mu _{\, \, \,  \nu} X^\nu(\sig=\pi) ,  \frac{1}{\pi} X'^\mu \) . 
\ee

We will choose to calculate the energy by integrating on the fixed $X^0(\tau, \sig)$ slice, because we want to  identify the time dependence of the mass as seen by a space-time observer:
\be \label{eq:charge}
\begin{split}
	{\cal E} &= \int_{X^0=\rm const} \( d\sig P^0_\tau + d\tau P^0_\sig \)
	= \int_0^\pi d\sig \left( P^0_\tau(X^0,\sig) +\frac{\de \tau(X^0,\sig)}{\de \sig} P^0_\sig(X^0,\sig) \right) \\
	&= \int d\sig \left( \frac{b^2 \cosh^2(\chi_0-\chi\sig) + \frac{(\pi\chi X^0)^2}{\cosh^2(\chi_0-\chi\sig)}}{\pi^2 \sqrt{b^2 \cosh^2(\chi_0-\chi\sig) + (\pi\chi X^0)^2}} \right) + \Big\{ e_0 E X^1(X^0,0) +e_\pi E X^1(X^0,\pi) \Big\}  \\
	& = {\cal E}_{\rm bulk} + {\cal E}_{\rm electric} \\
	&= \frac{b}{\pi}\sqrt{\frac{\chi}{E(e_0+e_\pi)}} \ .
\end{split}
\ee
The energy splits into two terms, where the term in curly braces is readily interpretable as the electric potential energy of the system. This allows us to identify the first term as the energy of the bulk of the string. The time dependence cancels between the two terms so that the total energy is constant.

In fact, the bulk energy ${\cal E}_{\rm bulk}$ can also be computed by integrating an effective mass density $T_0 \gamma_T$ along the string:
\be \label{eff mass}
	{\cal E}_{\rm bulk} = \int d \sigma T_0 \gamma_T \, , 
\ee
where $\gamma_T$ is the Lorentz factor for the component of the velocity that is transverse to the string, and $T_0$ is the string tension (equal to $1/\pi$ in our units). 
This is the same expression for the string energy that one would find in the $E=0$ case~\cite{zwiebach2009first}, as we will see in \appref{app:scatwork} in the scattering frame.

\section{Energy conservation in brane scattering} \label{app:scatwork}

Here we consider the energies of strings that are produced when branes scatter at constant velocity. For concreteness, consider a string stretched between two parallell $p-$branes, one at $X^2=b$ and moving in the $X^1$-direction with velocity $v=\tanh(\pi \chi)$, and one which is at rest at $X^2=0$. (Note that there is no loss of generality here, because for pair of scattering branes we can always boost to the frame where one is at rest.) One can find the classical solution for such a string by T-dualizing \eqref{eq:electricsol}:
\be
(X^0,X^1,X^2)=\left( \frac{b}{\pi \chi} \sinh(\chi \tau) \cosh(\chi \sig)  ,\:  \frac{b}{\pi \chi} \sinh(\chi \tau) \sinh(\chi \sig) ,\: \frac{b}{\pi}\sig \right),
\ee
or
\be
	X^1 = X^0 \tanh(\chi \sig) \, , \qquad 
	v = \partial_0 X^1 = \tanh(\chi \sig) \, , \qquad
	\gamma = \cosh(\sig \chi) \, .
\ee 

We would like to understand the energy of the string, as well as the analog of the electric potential energy: the work that the string does on the brane.  The force that the string endpoint exerts on the brane in its restframe is simply the usual string tension $T_0$ (which in the rest of the paper we have set to $1/\pi$).  In the frame where the brane moves with velocity $v$, the $X^1$-component of this force (the only component that does any work, since the brane's velocity is zero in the other directions) is:
\be
F_1= T_0 \frac{\pi \chi X^0}{\sqrt{b^2 \cosh^2(\chi \sig) +(\pi \chi X^0)^2}}= T_0 \frac{\pi \chi X^0}{\sqrt{b^2 \gamma^2 + (\pi \chi X^0)^2}}.
\ee
Define $\gamma_T$ as the Lorentz factor transverse to the string:
\be
\begin{split}
v_T &= v \cos(\theta) =  \frac{b \tanh{\chi \sig}}{\sqrt{b^2 +\frac{\chi^2 t^2}{\cosh^4(\chi \sig)}}}\\
\gamma_T &= \left( 1 - \frac{b^2 \tanh^2(\chi \sig)}{b^2 + \frac{\chi^2 t^2}{\cosh^4(\chi \sig)}} \right)^{-1/2} = \sqrt{ \frac{b^2 \cosh^4(\chi \sig) + \chi^2 t^2}{b^2 \cosh^2(\chi \sig) + \chi^2 t^2}}. \\
\end{split}
\ee
Now, the work is simply:
\be
\begin{split}
W &= \int dX^1 \: F_1(\sig = \pi) = \int_0^{X^0} dX^0 \: \tanh(\pi \chi)F_1(\sig = \pi) \\
&= T_0 \left( -\frac{b}{\pi \chi}\sinh(\pi \chi) + \frac{\tanh(\pi \chi)}{\pi \chi}\sqrt{b^2 \cosh^2(\pi \chi) + (\pi \chi X^0)^2}\right)
\end{split}
\ee

We have the work done on the brane, but in order to check energy conservation we also need to consider the energy of the string itself. This is the integral of the effective mass density along the length of the string:
\be \label{EFree}
\begin{split}
{\cal E}_{\rm bulk} &= \int dl \: T_0 \gamma_T = \int d\sig \: \sqrt{ \left(\partial_\sig X^1\right)^2 +\left(\frac{b}{\pi}\right)^2} \, T_0\, \gamma_T \\
&=T_0 \frac{\tanh(\pi \chi)}{\pi \chi}\sqrt{b^2 \cosh^2(\pi \chi) + (\pi \chi X^0)^2}.
\end{split}
\ee
It is important to note the asymptotic behavior of \eqref{EFree}: at late times (or small impact parameter) it is simply $T_0 v X^0 = T_0 l$.  

The conserved energy of the system is:
\be \label{stringen}
{\cal E}_{\rm bulk}  - W = T_0 \frac{b}{\pi \chi}\sinh(\pi \chi).
\ee
Energy is conserved, just as in the case of a constant electric field.  Note that this constant is arbitrary in both cases: in the case of the electric field it can be altered by changing the zero-point of the electric potential, and in this case it can be altered by measuring the work done on the brane starting from a different reference point.

\bibliographystyle{klebphys2}
\bibliography{Bibliography}{}

\end{document}